
\newskip\oneline \oneline=1em plus.3em minus.3em
\newskip\halfline \halfline=.5em plus .15em minus.15em
\newbox\sect
\newcount\eq
\newbox\lett

\def\simlt{\mathrel{\lower2.5pt\vbox{\lineskip=0pt\baselineskip=0pt
           \hbox{$<$}\hbox{$\sim$}}}}
\def\simgt{\mathrel{\lower2.5pt\vbox{\lineskip=0pt\baselineskip=0pt
           \hbox{$>$}\hbox{$\sim$}}}}

\newdimen\short
\def\adv{\global\advance\eq by1}
\def\set#1#2{\setbox#1=\hbox{#2}}
\def\nextlet#1{\global\advance\eq by-1\setbox
                \lett=\hbox{\rlap#1\phantom{a}}}

\newcount\eqncount
\eqncount=0
\def\equn{\global\advance\eqncount by1\eqno{(\the\eqncount)} }
\def\put#1{\global\edef#1{(\the\eqncount)}           }

\def\np{{\it Nucl. Phys.}}
\def\pl{{\it Phys. Lett.}}
\def\pr{{\it Phys. Rev.}}
\def\prl{{\it Phys. Rev. Lett.}}

\def\cGsw{1}
\def\cAbl{2}
\def\cAnt{3}
\def\cAqm{4}
\def\cAB{5}
\def\cKs{6}
\def\cK{7}
\def\FC{8}
\def\cDQZ{9}
\def\cMRS{10}
\def\ckf{11}
\def\CDF{12}
\def\LHC{13}
\def\cDlr{14}

\magnification=1200
\hsize=6.0 truein
\vsize=8.5 truein
\baselineskip 14pt

\nopagenumbers

\rightline{hep-ph/9403290}
\rightline{CPTH-A293.0294}
\rightline{IEM-FT-84/94}
\rightline{March 1994}
\vskip 1.0truecm
\centerline{\bf PRODUCTION OF KALUZA-KLEIN STATES}
\centerline{\bf AT}
\centerline{\bf FUTURE COLLIDERS}
\vskip 1.0truecm
\centerline{{\bf I. Antoniadis}, {\bf K. Benakli}}
\vskip .5truecm
\centerline{{\it Centre de Physique Th{\'e}orique}
\footnote{$^*$}{\it Laboratoire Propre du CNRS UPR A.0014}}
\centerline{\it Ecole Polytechnique, 91128 Palaiseau, France}
\vskip .5truecm
\centerline{{\bf M. Quir\'os}}
\vskip .5truecm
\centerline{{\it Instituto de Estructura de la Materia}}
\centerline{\it CSIC, Serrano 123, 28006 Madrid, Spain}

\vskip 2.5truecm
\centerline{\bf ABSTRACT}
\vskip .5truecm

Perturbative breaking of supersymmetry in four-dimensional string
theories predict in general the existence of new large dimensions at the
TeV scale. Such large dimensions lie in a domain of energies accessible to
particle accelerators. Their main signature is the production of
Kaluza-Klein excitations which can be detected at future colliders. We
study this possibility for hadron colliders (TEVATRON, LHC) and $e^+ e^-$
colliders (LEP-200, NLC-500).

\hfill\break
\vfill\eject

\footline={\hss\tenrm\folio\hss}\pageno=1

New degrees of freedom are required in any attempt of unification of the
electroweak and strong interactions with gravity. Among these attempts,
only superstring theory is known to provide a consistent quantum
theory of gravity [\cGsw]. Strings predict two kinds of new degrees of
freedom: (i) Superheavy oscillation modes whose characteristic mass scale
is given by the inverse of the string tension ${\alpha'}^{-1/2}\sim
10^{18}$ GeV. These states are important at very short distances of
order of the Planck length and modify the ultraviolet behavior of
gravitational interactions. (ii) States associated to the internal
compactified space whose presence is required from the fact that
superstring theory in flat space is anomaly free only in ten dimensions.
Usually, the size of this internal space is also made too small to give
any observable effect in particle accelerators.

In this work we are interested in the possibility of having one or two
large internal dimensions at a scale accessible to future experiments. The
presence of such large dimensions is motivated by superstring
theory with perturbative breaking of supersymmetry [\cAbl-\cAnt]. In this
context, their size is inversely proportional to the scale of supersymmetry
breaking which must be of order of the electroweak scale in order to
protect the gauge hierarchy. In contrast to field theoretical expectations,
string theory allows the existence of such large dimension(s) consistently
with perturbative unification of low-energy couplings in a class of models
based on orbifold compactifications [\cAnt].

Properties and implications of string models with perturbative
breaking of supersymmetry, were studied in the case of minimal
embedding of the standard model [\cAqm]. It was pointed out that the main
signature of the large extra dimension(s) in these constructions is the
appearance of a tower of excitations for the gauge bosons and higgses with
the same gauge quantum numbers.
These Kaluza-Klein (KK) states, characteristic of all theories with
compactified dimensions, have masses
$$
m_n^2 = m_0^2 + {{\vec n}^2 \over R^2}\ ,
\equn\put\kkmass
$$
where $R$ denotes the common radius of the $D$ large internal dimensions
($D$ = 1 or 2), while ${\vec n}$ is a $D$ dimensional vector with
integer entries. $m_0$ stands for $R$-independent contributions coming
from the electroweak symmetry breaking. In the following, $m_0$ will be
neglected since we are interested in values of $R^{-1}\simgt 200\ {\rm
GeV}$. Another characteristic of these models is that the massive KK-modes
are organized in multiplets of $N=4$ supersymmetry, which contain one
vector boson, four two-component fermions and three complex scalars. Since
KK-excitations exist only for gauge bosons and higgses, they form, in the
minimal case, the adjoint representation of $SU(3)\times SU(3)_c$ [\cAB].
In fact, the adjoint representation of the first $SU(3)$ contains in
addition to the adjoint of $SU(2)\times U(1)$ two doublets with the gauge
quantum numbers of the two higgses of the minimal supersymmetric
extension of the standard model (MSSM). In contrast, quarks
and leptons have no excitations.

All massive KK-states are unstable. They disintegrate into quarks
and leptons within a short lifetime, of the order of $10^{-26}$ seconds,
when the size of the compact dimension(s) is 1 TeV$^{-1}$ [\cAB].

Present experimental limits have been obtained from an analysis of
the effective four-fermion operators which arise from the exchange of the
massive KK-modes. Model independent bounds can be derived from the
modifications of known cross-sections generated by such effective
interactions [\cKs]. A computation of these operators in orbifold models
has shown that the present limits are $R^{-1}\simgt 185\ {\rm GeV}$ for one
large extra dimension, while
$R^{-1}\simgt 1.4\ {\rm TeV},1.1\ {\rm TeV},1\ {\rm TeV} $ for
two large dimensions in the case of ${\bf Z}_3$, ${\bf Z}_4$ and ${\bf
Z}_6$ orbifolds, respectively [\cAB]. These values leave open the
exciting possibility of detecting such extra dimensions at TEVATRON, or at
future colliders (LHC, LEP-200, NLC-500).
Below, we compute the cross sections and asymmetries when
KK-excitations are produced in these colliders.

\vskip 1.0cm
\centerline{\bf THE COUPLINGS OF KALUZA-KLEIN STATES}
\vskip 0.5cm

For the purpose of this work, we need the values for the couplings of
KK-modes to quarks and leptons. It is easy to show that, in the minimal
case, these couplings are non-vanishing only for vector excitations
with the quantum numbers of the standard model gauge bosons, and for
scalar excitations with the quantum numbers of the two Higgs doublets
[\cAB]. The former couple through gauge interactions, while the latter
through Yukawa interactions.

In the large radius limit, the tree level couplings of the massive
KK-modes to matter fermions are all equal to the corresponding coupling
of the lowest excitation, up to negligible $\alpha'/R^2$ corrections.
However, the computation of any low energy process requires the knowledge
of the effective renormalized couplings. In string theory, which is
ultraviolet finite, the low energy running of coupling constants can be
obtained from the logarithmic infrared divergences of the corresponding
on-shell scattering amplitudes [\cK]. This procedure is appropriate for
couplings of massless particles and it becomes less clear when external
legs are massive KK-modes.

One way to define the one loop renormalized
couplings of the vector KK-excitations to quarks and leptons is to
extrapolate the results of ref.[\cAB] for the coefficients of the
corresponding four-fermion effective operators. These coefficients were
computed in the kinematic region of external momenta smaller than the
compactification scale, by summing over all one loop infrared divergent
bubbles. In fact at the tree level, the four-fermion amplitude is
proportional to:
$$
{\cal A}(p,R)= {g^2\over p^2} + g^2\delta S\ ,
\equn\put\Sbar
$$
where $g$ is the tree level string coupling. The first term
in the r.h.s. of {\Sbar} denotes the contribution of the massless gauge
boson exchange (considering only the $s$ channel), while the second term
describes a contact interaction coming from the exchange of the massive
KK-modes [\cAB]. The renormalized one loop amplitude becomes
[\cAB] \footnote{$^1$}{In the simplest case where all matter fermions come
from the same fixed point of the orbifold.}:
$$
{\cal A}_{ren}(p,R)=
{g^2(p)\over p^2} + {g^4(p)\over g^4_U(p)} g^2\delta S\ ,
\equn\put\Srenf
$$
where $g(p)$ is the usual running coupling constant associated to
the massless gauge boson, while $g_U(p)$ is the effective coupling
which takes into account only the contribution to the $\beta$-function of
the massless states which have KK-excitations, {\it i.e.} gauge boson and
Higgs supermultiplets. Equation {\Srenf} suggests that the effective
coupling of the massive KK-modes can be identified as:
$$
\alpha^*(p)\equiv {g^{*2}(p) \over 4\pi} = {g^4(p) \over g^4_U(p)}
\alpha\ .
\equn\put\effcoupl
$$
where $\alpha\equiv {g^{2} \over 4\pi}$. In fact, though the expression
{\Srenf} has been obtained in the low momentum region ($p R \ll 1$), the
coefficients of the four-fermion operators ($\sim g^{*2}(p)
/m_n^2(p)$) in the effective low-energy theory should not run for scales
$p^2 R^2\simlt 1$ \footnote{$^2$}{A direct calculation of the
electromagnetic corrections to muon beta decay shows that this is
the case in the standard model of electroweak interactions
[\FC]. More precisely,
the Fermi constant $G_F \sim g^2(p)/M_W^2(p)$ does not run
(it is only finitely renormalized).}. For scales $p^2 R^2 >1$ the
coupling $g^*(p)$ should run with the scale as in an ordinary field
theory. However to compute the running of $g^*(p)$ between
$1/R$ and $M_{GUT}$ we should properly take into account the
(unknown) threshold
effects due to the exceedingly large number of KK-excitations
with masses $m_n$. Fortunately, we only need in our calculation
the couplings $g^*(p)$ at scales $p^2 \le m_n^2$, where
the value of $g^*(p)$ is provided by the couplings
$g(p)$ and $g_U(p)$, which have a well known running since they
are renormalized only by the states
belonging to the usual massless sectors of the theory.

Using the values for the standard model coupling constants
$g_2(M_Z) = 0.656$ and $g'(M_Z) = 0.357$ and the particle
content of the  MSSM ({\it i.e.} gauge coupling unification at
$M_{GUT} \sim 2 \times 10^{16}$ GeV, with $\alpha_{GUT}\sim 1/25$)
one finds that in the region of energies of a few
hundred GeV to a few TeV, which will be explored in future colliders,
the effective coupling of the massive $SU(2)$
($\vec{W}_n^*$) excitations $g^*_2(p)$
is of the order of $10^{-2}$ and can be neglected. On the other hand, the
value of the effective coupling of the massive $U(1)_Y$
($B_n^*$) excitations $g'^*(p)$ varies in the range 0.25-0.27\ .

We should stress here that the quoted values of $g_2^*(p)$ and
$g'^*(p)$ are sensitive to the unification scale. In fact, in models
where $M_{GUT}$ is shifted to $\sim 10^{18}$ GeV, $g_2^*(p)$ is no longer
negligible as compared to $g'^*(p)$. In those cases, $\gamma_n^*$ and
$Z_n^*$ have sizeable couplings to fermions and should both be considered
in the production cross sections and asymmetries computed below,
therefore modifying correspondingly the bounds on $R^{-1}$.

\vskip 1.0cm
\centerline{\bf PRODUCTION AT HADRON COLLIDERS}
\vskip 0.5cm

Among the various KK-excitations of different spins, the easiest to detect
at future colliders are the vectors with the quantum numbers of the
electroweak $SU(2)\times U(1)_Y$ gauge bosons. Here, we shall follow the
method of ref.[\cDQZ], where a similar analysis was performed for the
production, at hadron colliders, of $Z'$ vector bosons present in $E_6$
superstring-inspired models. It was pointed out that the most efficient
way of observing new gauge bosons in proton-(anti)proton collisions is to
identify charged leptons $l^{\pm}$ in the final state. Thus, we need to
compute the cross-sections for the Drell-Yan processes $pp \rightarrow
l^+l^-X$ at LHC, or $p{\bar p} \rightarrow l^+l^-X$ at TEVATRON, with
$l=e,\mu,\tau$. These processes receive also a contribution from the
exchange of scalars with quantum numbers of the higgses, which is
however suppressed by the small Yukawa couplings. Scalars, as well as the
remaining vector bosons, can be produced through other processes and they
are in general harder to detect.

As pointed out above, because of the accidental suppression of the
effective coupling of the massive $SU(2)$ vector excitations, only the
$U(1)_Y$ excitations $B^*_n$ ($n\ne 0$) couple to leptons and contribute
to Drell-Yan processes. These states have masses given in eq.{\kkmass} and
they couple to matter fermions through the effective interaction:
$$
g'^*(p){\bar \psi^k}{\gamma_\mu}(v_k + a_k \gamma_5){\psi^k} B^{*\mu}_n\ ,
\equn\put\Bint
$$
where $k$ labels the different species of fermions, and $g'^*(p)$ is given
in eq.{\effcoupl}. The values of the vector couplings $v_k$ and the axial
couplings $a_k$ are given in table 1.

The interactions {\Bint} lead to rates of $B^*_n$ decays into fermions:
$$
\Gamma (B^*_n \rightarrow f{\bar f})= {g'^*}^2(m_n)\ {m_n\over 12\pi}C_f
(v_f^2 + a_f^2)\ ,
\equn\put\gf
$$
while the corresponding interactions with their scalar superpartners
$\tilde{f}_{R,L}$, lead to the decay rates:
$$
\Gamma (B^*_n\rightarrow{\tilde f}_{R\atop L}{\tilde{\bar{f}}}_{R\atop L})
= {g'^*}^2(m_n)\ {m_n\over 48\pi} C_f (v_f \pm a_f)^2 \ ,
\equn\put\gs
$$
where $C_f=1$ or 3 for color singlets or triplets, respectively. In the
above computation we neglected the masses of the fermions and of their
superpartners. The latter could be important for the lowest excitations,
but they are model dependent. In this approximation, where both the
supersymmetry and electroweak gauge symmetry breaking are neglected, the
width $\Gamma_n$ of the $B^*_n$ resonances is obtained by summing only over
the partial widths {\gf} and {\gs}. In fact, the decay rates of $B^*_n$
to a pair of states with gauge quantum numbers of the higgses vanish in
this limit, due to the conservation of the momentum associated to the
compactified dimension(s). The total width is then:
$$
\Gamma_n= {5\over 8\pi}{g'^*}^2 m_n\ .
\equn\put\gt
$$
Note that if one takes into account only the decay rates into fermions,
the width {\gt} is decreased by a factor $2/3$.

The proton-(anti)proton system has a center of mass energy
$\sqrt s = 16\ {\rm TeV}$ and $1.8\ {\rm TeV}$ for LHC and TEVATRON,
respectively. In the corresponding Drell-Yan process, the lepton pairs are
produced via the subprocess $q{\bar q}\rightarrow l^+ l^- X$ of center of
mass energy $M$. The two colliding partons take a fraction
$$
x_a={M \over \sqrt s}\ e^{y} \quad{\rm and}\quad
x_b={M \over \sqrt s}\ e^{-y}
\equn\put\momfrac
$$
of the momentum of the initial proton ($a$) and (anti)proton ($b$), with a
probability described by the quark or antiquark distribution functions
$f^{(a)}_{q,\bar q}(x_{a}, M^2)$ and $f^{(b)}_{q,\bar q}(x_{b}, M^2)$.
In our numerical computations, we will use the parton distribution
functions given by Martin, Roberts and Stirling (set $S_o$) [\cMRS].

The total cross section, due to the production of the KK-excitations
$B_n^*$, is given by:
$$
\sigma= \sum_{q={\rm quarks}} \int^{\sqrt s }_0 dM \int^{\ln (\sqrt s
/M)}_{\ln (M/\sqrt s)}dy \ g_q (y, M) S_q (y, M) \ ,
\equn\put\sig
$$
where
$$
g_q (y, M)= {M \over 18\pi} x_a x_b \ [f^{(a)}_q (x_a,M^2)
f^{(b)}_{\bar q} (x_b, M^2) + f^{(a)}_{\bar q} (x_a, M^2) f^{(b)}_q (x_b,
M^2)]\ ,
\equn\put\gq
$$
and
$$
S_q (y, M)= {g'^*}^4 {1\over N} \sum_{|{\vec n}| < R \sqrt s}{(v_q^2 +
a_q^2)(v_l^2 +a_l^2)
\over (M^2 -m^2_n)^2 + \Gamma^2_n m^2_n } \ ,
\equn\put\Sq
$$
where the factor $1/N$ comes from the ${\bf Z}_N$ orbifold projection
which identifies the state $|n>$ with $|\theta n>$, where
$\theta=e^{2i\pi\over N}$. In the case of two dimensions, we use the
complex notation $n=n_1+\theta n_2$ to represent the vector ${\vec
n}=(n_1,n_2)$, and ${\vec n}^2=|n|^2$. For the numerical results, we
restrict our analysis to the cases of one-dimensional ${\bf Z}_2$ and
two-dimensional ${\bf Z}_4$ orbifolds. In eq.{\Sq}, we have dropped the
interference terms which are negligible in our case, since the
KK-resonances do not overlap, as their widths {\gt} are small compared to
their relative distance due to the coupling constant suppression. This
property also allows to replace in eq.{\sig} the integration over $M$ by
the approximate expression
${\pi\over 2} \Gamma_n {d\sigma\over dM}{\mid_{M=m_n}}$ for each resonance.

The cross-section {\sig} has to be corrected by incorporating the
radiative corrections in the initial states. These corrections are usually
described by a multiplicative factor $K$ [\ckf]. We found $K\sim 1.3$ for
TEVATRON and $K\sim 1.1$ for LHC. After multiplication by $K$, the total
cross-sections for the production of $l^+ l^-$ pairs at TEVATRON and LHC
are plotted as a function of $R^{-1}$ in figs. 1 and 2, respectively.
In each figure, we present two curves for the cases of
one (solid lines) and two (dotted lines) large
compactified dimensions. The corresponding numbers of events are obtained
after multiplication by the integrated luminosity.

At TEVATRON, the CDF collaboration has collected an integrated luminosity
$\int {\cal L}dt= 21.4\ pb^{-1}$ during the 1992-93 running period. From
the non-existence of candidate events at $e^+e^-$ invariant mass above
350 GeV, and using the fact that the detection efficiency is rather flat
and $\sim$ 40 \% in the region between 350 GeV and 700 GeV
\footnote{$^3$}{We thank K. Maeshima for private communication on this
point.}, they obtain the bound [\CDF]
$$
\sigma(B_n)\cdot B(B_n\rightarrow e^+e^-)< 0.35\ pb\
(95\ \% \ C.L.)
\equn\put\bb
$$
{}From eq.{\bb} and the result in fig. 1 one deduces the bound
$R^{-1}\simlt 510\ {\rm GeV}$. We can scale the bound
{\bb} for the mass region below 700 GeV,
to the increased value
of luminosity accessible in the future, $\int {\cal L}dt\sim 100\ pb^{-1}$
$$
\sigma(B_n)\cdot B(B_n\rightarrow e^+e^-)\simlt 0.075\ pb\ (95\ \% \ C.L.)
\equn\put\cc
$$
which would translate in the (future) bound
$R^{-1}\simlt 650\ {\rm GeV}$. In particular,
this result implies that (in view of the limit of 1.1 TeV
from the analysis of effective four- fermion operators)
the KK-excitations in the case of {\bf Z}$_4$ orbifold
should not be detected at any future TEVATRON run. However, since those
limits are milder for {\bf Z}$_2$ orbifolds, the corresponding
KK-excitations could be produced
if they are lighter than $\sim$700 GeV. For LHC
(fig.2), using a luminosity of $\int {\cal L}dt\sim 10^5\  pb^{-1}$,
and an efficiency of $\sim$ 15 \% (which would amount at 95 \% C.L. to
a discovery limit of 20 events [\LHC]), one would obtain a bound
$$
\sigma(B_n)\cdot B(B_n\rightarrow e^+e^-)\simlt 2\times 10^{-4}\ pb\
(95\ \% \ C.L.)
\equn\put\cc
$$
{\it i.e.}, $R^{-1}\simlt 4.5\ {\rm TeV}$.

Above, we gave the results based on total cross sections. In the presence
of enough statistics one could in principle identify different resonances,
regularly spaced, by plotting the number of events as a function of the
lepton pair invariant mass $M$. This would be a clear signal for the
existence of new dimension(s). Of course, an identification of precise
couplings is also required but this is much harder to achieve at hadron
colliders.

\vskip 1.0cm
\centerline{\bf PRODUCTION AT $e^+e^-$ COLLIDERS}
\vskip 0.5cm

In contrast to the case of hadrons, in $e^+e^-$ colliders the invariant
mass of the produced fermion pairs is fixed (to a first approximation) and
the presence of $B_n^*$ resonances cannot be directly observed, unless
the machine energy happens to be very close to the mass of one of
the excitations, or else a scanning of energies is made. Moreover, since the
presently planned energies of next linear colliders (NLC-500) are small
compared to LHC, direct production of heavy vector bosons is more limited.
However, $e^+e^-$ experiments have a clean environment which allows to
perform high precision measurements. These could reveal the existence of
extra gauge bosons, through their virtual effects. The strategy is then to
compare accurate measured quantities, such as the total cross section
$\sigma_T$, the forward-backward asymmetry $A_{FB}$, and the ratio of
hadron to lepton production, to the values predicted by the standard
model. A possible disagreement could be interpreted as a signal for
new physics at a higher scale. Below, we compute the
deviations in $\sigma_T$ and $A_{FB}$ due to the exchange
of KK-modes, at LEP-200 and NLC-500.

The total cross section for the annihilation of unpolarized
electron-positron pairs $e^+e^-$, with a center of mass energy $\sqrt s$,
to lepton pairs  $l^+l^-$, through the exchange of vector
bosons in the s-channel, is given by:
$$
\sigma_T^0(s)={s\over 12 \pi} \sum_{\alpha ,\beta=\gamma, Z, B^*_n}g^2_{\alpha}
({\sqrt s}) g^2_{\beta} ({\sqrt s}) {(v^{\alpha}_e v^{\beta}_e+
a^{\alpha}_e a^{\beta}_e)(v^{\alpha}_l v^{\beta}_l + a^{\alpha}_l
a^{\beta}_l) \over (s -m^2_{\alpha} + i\Gamma{_\alpha}
m_{\alpha})(s-m^2_{\beta} - i\Gamma_{\beta} m_{\beta})} \ ,
\equn\put\eesig
$$
where the labels $\alpha ,\beta$ stand for the different neutral vector
bosons $\gamma$, $Z$, and $B^*_n$ with coupling constants
$g_{\alpha} = e$, $e/(\sin{\theta_w}\cos{\theta_w})$, and $g'^*$,
respectively; $\theta_w$ is the weak mixing angle. With the same notation,
the cross section for the forward-backward asymmetry is:
$$
\sigma_{FB}^0(s)={s\over 16 \pi} \sum_{\alpha ,\beta=\gamma, Z, B^*_n}
g^2_{\alpha} ({\sqrt s}) g^2_{\beta} ({\sqrt s}) {(v^{\alpha}_e
a^{\beta}_e+ a^{\alpha}_e v^{\beta}_e)(v^{\alpha}_l a^{\beta}_l +
a^{\alpha}_l v^{\beta}_l) \over (s -m^2_{\alpha} + i\Gamma_{\alpha}
m_{\alpha} )(s -m^2_{\beta} - i\Gamma_{\beta} m_{\beta})} \ .
\equn\put\eefb
$$

As we expect in experiments to be held at energies far from the
resonances peaks, the contribution of $B^*_n$ exchanges to the cross
sections {\eesig} and {\eefb} will be dominated by the $\gamma\ B^*_n$
interference terms, due to the mass suppression from the propagators.
These effects are small, and they must be computed with high
accuracy. It is then necessary to include radiative corrections, and
in particular the bremsstrahlung effects on the initial electron and
positron [\cDlr]. These are described by the convolution of {\eesig} and
{\eefb} with radiator-functions which describe the probability of having a
fractional energy loss, $x$, due to the initial state radiation:
$$
\sigma_T(s) =\int^{\Delta}_0 dx \sigma_T^0(s') r_T(x) \ ,
\equn\put\eesigr
$$
and
$$
A_{FB}(s) ={1\over \sigma_T} \int^{\Delta }_0 dx \sigma_{FB}^0(s')
r_{FB}T(x)\ ,
\equn\put\eefbr
$$
with
$$
s'=s(1-x)\ .
$$
In the above equations, $\Delta$ represents an experimental cut for
the energy of emitted soft photons in bremsstrahlung processes. The
radiator functions are given by [\cDlr]:
$$
\eqalign{
r_T(x) &=(1 + X) y x^{y -1}+ H_T(x)\cr
r_{FB}(x) &=(1 + X) y x^{y -1}+ H_{FB}(x)\ ,\cr}
\equn\put\rt
$$
with:
$$
\eqalign{
X &= {e^2(\sqrt{s}) \over 4\pi^2} [{\pi^2\over 3}- {1\over
2}+{3\over 2}(\log{{s\over m_e^2}}-1)]\cr
y &={2e^2(\sqrt{s}) \over 4\pi^2} (\log{{s\over m_e^2}}-1)\cr
H_T &= {e^2(\sqrt{s}) \over 4\pi^2} [{1+(1-x)^2\over x}(\log{{s\over
m_e^2}}-1)]-{y\over x}\cr
H_{FB} &= {e^2(\sqrt{s}) \over 4\pi^2} [{1+(1-x)^2\over x}{1-x\over
(1-{x \over 2})^2}(\log{{s\over m_e^2}}-1-\log{{1-x\over
(1-{x \over 2})^2}})]-{y\over x}\ ,\cr}
\equn\put\hf
$$
where $m_e$ is the electron mass.

We use ${\sqrt s}=190$ GeV for LEP-200 and 500 GeV for NLC-500, together
with the numerical values for the experimental cuts [\cDlr]:
$$
\Delta( {\rm LEP-200})= 0.770\quad {\rm and}\quad
\Delta( {\rm NLC-500})= 0.967\ ,
\equn\put\delt
$$
coming from the condition of removing the $Z$ boson tail, which
amounts to imposing the cut $s'\ge M_Z^2$.

In fig.3 we plot, for LEP-200, the ratios
$$
R_T=\left|{\sigma_T(s)-\sigma_T^{\rm SM}(s) \over \sigma_T^{\rm SM}(s)}
\right|
\equn\put\rt
$$
and
$$
R_{FB}=\left|{A_{FB}(s)-A_{FB}^{\rm SM}(s) \over A_{FB}^{\rm SM}(s)}
\right|,
\equn\put\rfb
$$
where $\sigma_T^{\rm SM}(s)$ and $A_{FB}^{\rm SM}(s)$ are the standard model
predictions for the total cross section and forward-backward asymmetry,
respectively. The cases of ${\bf Z}_2$ and ${\bf Z}_4$ orbifolds are
indicated with solid and dotted lines, respectively. The same plot for
NLC-500 is exhibited on fig.4.

We can see from figs. 3 and 4 that the measurement of either of these
quantities, $\sigma_T(s)$ and $A_{FB}(s)$ with a certain degree of
accuracy will translate into a lower bound on $R^{-1}$. For instance,
measuring both of them at LEP-200 with an error 1\% would
translate into a bound $R^{-1} \simgt 1.6$ TeV for the {\bf Z}$_2$
orbifold, and $R^{-1} \simgt 3.0$ TeV for the {\bf Z}$_4$
orbifold. At NLC-500 the bounds would become
$R^{-1} \simgt 4.9$ TeV for the {\bf Z}$_2$
orbifold, and $R^{-1} \simgt 9.2$ TeV for the {\bf Z}$_4$
orbifold.

\vskip.5cm

One of the reasons for building future colliders is the search for sparticles
which should be seen if supersymmetry is broken at low energy. This work shows
that the planned energies will also allow to partially investigate the origin
of the breaking. In particular, the cross sections derived above make
it clear that
LHC and NLC are able to test the possibility for perturbative breaking using
large extra dimension(s). Moreover, the possible detection of KK-states is very
exciting since such dimensions can be implemented consistently only in
superstring theory and, thus, provide a window to study a part of the massive
string spectrum at low energies.

\vskip 1.0cm
\noindent{\bf Acknowledgements}
\vskip.5cm
One of us (MQ) wishes to thank J. Benlloch and K. Maeshima for
useful informations concerning CDF results.
This work was supported in part by IN2P3-CICYT under contract PTh91-2,
in part by the EEC contracts SC1-CT92-0792 and CHRX-CT92-0004,
and in part by the CICYT contract AEN90-0139.

\vskip 1.5cm
\centerline{\bf REFERENCES}
\vskip 0.5cm

\parskip=-3 pt

\item{[{\cGsw}]} See for example M.B. Green, J.H. Schwarz and E. Witten,
{\it Superstring theory}, Cambridge Press, 1987.\hfill\break

\item{[{\cAbl}]} T. Banks and L. Dixon, {\np} {\bf B307} (1988) 93; I.
Antoniadis, C. Bachas, D. Lewellen and T. Tomaras, {\pl} {\bf 207B}
(1988) 441; C. Kounnas and M. Porrati, {\np} {\bf B310} (1988) 355; S.
Ferrara, C. Kounnas, M. Porrati and F. Zwirner, {\np} {\bf B318} (1989)
75.\hfill\break

\item{[{\cAnt}]} I. Antoniadis, {\pl} {\bf 246B} (1990) 377; Proc.
PASCOS-91 Symposium, Boston 1991 (World Scientific, Singapore)
p.718.\hfill\break

\item{[{\cAqm}]} I. Antoniadis, C. Mu\~noz and M. Quir\'os, {\np} {\bf
B397} (1993) 515.\hfill\break

\item{[{\cAB}]} I. Antoniadis and K. Benakli, preprint {\bf
CPTH-A0793} (1993) .\hfill\break

\item{[{\cKs}]} V.A. Kostelecky and S. Samuel, {\pl} {\bf270B} (1991) 21.
\hfill\break

\item{[{\cK}]} V.S. Kaplunovsky, {\np} {\bf B307} (1988) 145 and {\it
Errata} STANFORD-ITP-838 preprint (1992).\hfill\break

\item{[{\FC}]} W. Marciano, {\pr} {\bf D20} (1979) 274;
F. Antonelli and L. Maiani, {\np} {\bf B186} (1981) 269.\hfill\break

\item{[{\cDQZ}]} F. Del Aguila, M. Quir\'os and F. Zwirner, {\np} {\bf
B287} (1987) 419.\hfill\break

\item{[{\cMRS}]} A.D. Martin, R.G. Roberts and W.J. Stirling, {\pr}
{\bf D47} (1993) 867.\hfill\break

\item{[{\ckf}]} R. Hamberg, W.L. van Neerven and T.Matsuura,
{\np} {\bf B359} (1991) 343.\hfill\break

\item{[{\CDF}]} CDF Collaboration, F. Abe {\it et al.}, {\prl} {\bf 68}
(1992) 1463; K. Maeshima, CDF Collaboration,
preprint Fermilab-Conf-94/063-E, to appear in the {\it Proceedings of
9th Topical Workshop on
$\bar{p}p$ Collider Physics}, Tsukuba, Japan (October 1993). \hfill\break

\item{[{\LHC}]} See, e.g.: F. Pauss, in the Large Hadron Collider
Workshop, Aachen, October 1990; preprint CERN 90-10 and ECFA 90-133,
Vol. I, Page 118. \hfill\break

\item{[{\cDlr}]} A. Djouadi, A. Leike, T. Riemann, D. Schaile and C.
Verzegnassi, {\it Z. Phys.} {\bf C56} (1992) 289.\hfill\break

\vfill\eject

\vbox {\tabskip=0pt \offinterlineskip\def\tablerule{\noalign{\hrule}}
\def\tv{\vrule height 20pt depth 5pt}\halign to 13cm {\tabskip=0pt
plus 20mm \tv\hfill\quad#\qquad\hfill &\tv\hfill\quad#\qquad\hfill
&\tv\hfill\quad# \qquad\hfill &\tv\hfill\quad#\quad\hfill
&\tv#\tabskip=0pt\cr\tablerule  Coupling & photon &
$Z$ &
  $B_n$ &\cr\tablerule
$v_u$&2/3&${1\over 4}-{2\over 3}s_w^2$ &$-{5\over 12}$
&\tabskip=0pt\cr $a_u$&0&$-{1\over 4}$&$-{1\over 4}$ &\tabskip=0pt\cr
$v_d$&-1/3&$-{1\over 4}+{1\over 3}s_w^2$&${1\over 12}$ &\tabskip=0pt\cr
$a_d$&0&${1\over 4}$&${1\over 4}$ &\tabskip=0pt\cr
$v_e$&-1&$-{1\over 4}+s_w^2$&${3\over 4}$ &\tabskip=0pt\cr
$a_e$&0&${1\over 4}$&${1\over 4}$ &\tabskip=0pt\cr
$v_\nu$&0&${1\over 4}$&${1\over 4}$ &\tabskip=0pt\cr
$a_\nu$&0&$-{1\over 4}$&$-{1\over 4}$ &\tabskip=0pt\cr\tablerule}}

\vskip 1.0truecm
\centerline{\bf Table 1.}
\vskip 0.5truecm
Couplings of the matter particles to the standard gauge bosons and
the $B_n$ KK-excitations (we used
$s_w\equiv \sin (\theta_w)$ ).

\vfill\eject

\centerline{\bf Figure Captions}
\vskip 0.5truecm

\item{\bf Fig. 1} Total cross-section $\sigma (p \bar p \rightarrow l^+
l^- X) $ at Tevatron as a function of the compactification scale
$R^{-1}$ (solid and dotted lines correspond to the one-dimensional
{\bf $Z_2$} and two-dimensional {\bf $Z_4$} orbifolds, respectively).
\hfill\break

\item{\bf Fig. 2} Total cross-section $\sigma (p p \rightarrow l^+
l^- X) $ at LHC as a function of the compactification scale
$R^{-1}$ (solid and dotted lines correspond to the one-dimensional
{\bf $Z_2$} and two-dimensional {\bf $Z_4$} orbifolds, respectively).
\hfill\break

\item{\bf Fig. 3} Expected deviations from the standard model for the
total and forward - backward asymmetry cross-sections at LEP-200, as a
function of the compactification scale  $R^{-1}$ (solid and dotted
lines correspond to the one-dimensional {\bf $Z_2$} and
two-dimensional {\bf $Z_4$} orbifolds, respectively).\hfill\break

\item{\bf Fig. 4} Expected deviations from the standard model for the
total and forward - backward asymmetry cross-sections at NLC-500 as a
function of the compactification scale $R^{-1}$ (solid and dotted
lines correspond to the one-dimensional {\bf $Z_2$} and
two-dimensional {\bf $Z_4$} orbifolds, respectively).\hfill\break

\end